%% file: QIS.tex
\documentclass[a4paper,oneside,11pt]{article}
\usepackage{epsfig,multicol}
\usepackage{graphicx}
\usepackage{amsmath,amssymb}
\usepackage{mathrsfs}

\begin{document}
\setcounter{page}{1}

\begin{center}
{\Large \bf Effective potential for relativistic scattering}
\end{center}

\vskip 19.6ex

\begin{center}
Mahmut Elbistan$^1$\footnote{\tt elbistan@impcas.ac.cn}\,,\ \  
Pengming Zhang$^1$\footnote{\tt zhpm@impcas.ac.cn} 
\ \ and\ \ 
J\'anos Balog$^{1,2}$\footnote{\tt balog.janos@wigner.mta.hu}
\\
\vskip 3ex
$^1$ {\it Institute of Modern Physics, 
Chinese Academy of Sciences,\\ Lanzhou 730000, China}\\
\vskip 1ex
$^2${\it MTA Lend\"{u}let Holographic QFT Group, Wigner Research Centre} \\
{\it H-1525 Budapest 114, P.O.B. 49, Hungary}\\

\end{center}


\vskip 14.5ex

\noindent
We consider quantum inverse scattering with singular potentials 
and calculate the Sine-Gordon model effective potential in the laboratory 
and centre-of-mass frames. The effective potentials are frame dependent but
closely resemble the zero-momentum potential of the equivalent 
Ruijsenaars-Schneider model.

\newpage


\newcommand{\beq}{\begin{eqnarray}}
\newcommand{\eeq}{\end{eqnarray}}
\newcommand{\ee}{\end{equation}}



\section{Introduction and motivation}

\input QISmotiv

\section{Effective potentials}

\input effpot

\section{Quantum inverse scattering with singular potentials}

\input singular

\section{Sine-Gordon effective potential in the laboratory frame}

\input LAB

\section{Sine-Gordon effective potential in the centre of mass frame}

\input COM

\section{Summary and conclusion}

\input QISconcl

\vspace{5ex}
\begin{center}
{\large\bf Acknowledgments}
\end{center}

This work was supported by the Major State Basic Research Development 
Program in China (No. 2015CB856903), 
the National Natural Science Foundation of China (Grant No. 11575254) and
by the Hungarian National Science Fund OTKA (under K116505). 
J.~B. would like to thank the CAS
Institute of Modern Physics, Lanzhou, where most of this work has been 
carried out, for hospitality.

\par\bigskip


\appendix

\section{Scattering and inverse scattering for the $1/\sinh^2x$ potential}

\input sinh

\section{The Sine-Gordon S-matrix}

\input SGSmatrix

\section{Determinant solution}

\input detsol

\input{bibliography}

\end{document}

%% file: QISmotiv
In recent years advances in lattice QCD techniques made possible to 
measure and study forces between nucleons. A major success was the first
principles calculation of the two-nucleon potential by the HAL QCD
collaboration \cite{Ishii:2006ec,Aoki:2008hh,Aoki:2009ji},
which was later extended to nucleon-hyperon
interactions \cite{Inoue:2010hs,Inoue:2010es}
and to the study of three-baryon forces \cite{Doi:2011gq}. Three-neutron
(and higher) interactions are crucial to determine the correct nuclear 
equation of state, which is used in the calculation of the mass and radius 
of neutron stars. Gravitational wave signals expected from inspiraling neutron 
star systems are sensitive to the resulting mass-radius relation.

The HAL QCD method \cite{Ishii:2006ec} is based on measuring the 
Nambu-Bethe-Salpeter (NBS)
wave function $\Psi_E({\mathbf x})$ of a two-nucleon state which satisfies
(in the centre of mass frame) the \lq\lq Schr\"odinger equation"
\begin{equation}
\left[-\frac{1}{m}\nabla^2+U_E({\mathbf x})\right]\Psi_E({\mathbf x})=
E\,\Psi_E({\mathbf x}),
\end{equation}
where $m$ is the nucleon mass. Due to the relativistic nature of the problem, the
NBS \lq\lq potential" $U_E({\mathbf x})$ is energy-dependent. This energy-dependence
however is found to be weak and the NBS potential at low energies resembles the
phenomenological nuclear potential used in nuclear physics for many 
decades \cite{NuPo1,NuPo2,NuPo3}. In particular, at short distances it has 
a characteristic repulsive core.

The problem of energy dependence can be studied in some $1+1$ dimensional integrable
models \cite{Aoki:2008yw}. 
The Ising model and the O$(3)$ nonlinear $\sigma$-model were studied
and it was found that at low energies the energy-dependent $U_E({\mathbf x})$ can
be well approximated by its zero-momentum limit (corresponding to the case where the
relative momentum of the two-particle state vanishes). The problem was also studied
in the Sine-Gordon (SG) model \cite{BZ}. In the semiclassical limit an 
energy-independent effective potential was constructed, which exactly reproduces 
the semiclassical time delays for all energies. This could be compared to the
zero-momentum potential, which is explicitly known in this model from its
equivalent Ruijsenaars-Schneider (RS) formulation \cite{RS1,RS2}.

In this paper we continue to study the notion of effective potential in the 
integrable (analytically solvable) SG model in $1+1$ dimension. We model the way
the phenomenological potential was determined from scattering experiments: we
require that the quantum mechanical effective potential exactly reproduces the
(analytically known) scattering phase shifts at all energies. The price we have
to pay is that the effective potential is frame dependent. We will construct the
effective potential in the laboratory frame of the scattering process and also in
the centre of mass frame of the two particles. We will compare them to each other
and to the zero-momentum potential known from the RS formulation of the model.

The paper is organized as follows. In section 2 we define the notion of effective
potential for relativistic scattering. Section 3 is a review of quantum mechanical
inverse scattering in one dimension. We generalize known results for the case of
singular potentials. In section 4 and 5 we calculate the effective potential for
soliton-soliton scattering in the SG model in the laboratory and centre-of-mass
frames, respectively. Section 6 is a short summary of the results and contains our
conclusions. Some technical details and examples can be found in the appendices
together with the summary of the scattering phase shifts in the SG model.

%% file: effpot
We will study the one-dimensional scattering of two identical partices
of mass $m$ (with positions $x_1$, $x_2$ and momenta $p_1$, $p_2$), whose 
interaction has a strong repulsive core which does not allow the particles
to come close to each other. If initially particle 1 is to the left of 
particle 2 then $x_2>x_1$ at all times. Initially $p_1>p_2$: 

\hspace{2.5cm}$\bullet$ { $\longrightarrow$} $p_1$
\hspace{6.0cm} $p_2$ { $\longleftarrow$} $\bullet$

\hspace{2.4cm}$x_1$\hspace{8.8cm}$x_2$

Asymptotically, for $(x_2-x_1)\to\infty$, the 2-particle wave function 
$\Phi(x_1,x_2)$ is a superposition of free waves:
\beq
\Phi(x_1,x_2)\approx \Phi_{\rm as}(x_1,x_2)=
{\rm e}^{i(k_1x_1+k_2x_2)}+S(p_1,p_2){\rm e}^{i(k_2x_1+k_1x_2)},
\qquad\quad x_2-x_1>>0.
\label{asPhi}
\eeq 
Here the first term is the incoming free wave and the second one is the
outgoing free wave which has picked up the phase factor $S(p_1,p_2)$
as a result of the interaction. We have introduced the wave vectors
$k_j=p_j/\hbar$, $j=1,2$.

For relativistic scattering, the \lq\lq S-matrix" $S(p_1,p_2)$ is a function
of the relative rapidity of the particles:
\beq
S_{\rm R}(p_1,p_2)
=-\Sigma(\theta_1-\theta_2),\qquad\qquad p_j=mc\sinh\theta_j.
\eeq 

For non-relativistic scattering we can use a quantum-mechanical description
with a potential depending on the relative distance of the particles. The
Hamilton operator has the form
\beq
\hat{\cal H}=-\frac{\hbar^2}{2m}\,\frac{\partial^2}{\partial x_1^2}
-\frac{\hbar^2}{2m}\,\frac{\partial^2}{\partial x_2^2}+U(x_2-x_1).
\eeq 
We have to find a solution of the Schr\"odinger equation
\beq
\hat{\cal H}\Phi=E\Phi,\qquad\qquad E=\frac{\hbar^2}{2m}(k_1^2+k_2^2)
\eeq 
with asymptotics (\ref{asPhi}). Separating the centre of mass and relative 
motions we can write
\beq
\Phi(x_1,x_2)={\rm e}^{iK(x_1+x_2)}\Psi(x_2-x_1),
\eeq 
where the relative wave function satisfies the Schr\"odinger equation
\beq
-\frac{\hbar^2}{m}\Psi^{\prime\prime}(x)+U(x)\Psi(x)=\frac{\hbar^2}{m}
\kappa^2\Psi(x).
\eeq 
Here
\beq
k_1=K+\kappa,\quad k_2=K-\kappa,\qquad\quad
 E=\frac{\hbar^2}{m}(K^2+\kappa^2),\qquad \kappa>0.
\eeq 
The $x\to\infty$ asymptotics of the relative wave function is required to be of the
form
\beq
\Psi(x)\approx \Psi_{\rm as}(x)=
-{\cal T}(\kappa){\rm e}^{i\kappa x}+{\rm e}^{-i\kappa x},
\qquad x>>0.
\label{asPsi}
\eeq 
Comparing to (\ref{asPhi}) gives
\beq
S_{\rm NR}(p_1,p_2)=-{\cal T}\left(\frac{p_1-p_2}{2\hbar}\right).
\eeq 
We can simplify the problem by introducing a length scale $\ell$ and rescaling 
the variables. We introduce
\beq
u(x)=\Psi(\ell x),
\eeq 
which satisfies
\beq
-u^{\prime\prime}(x)+q(x)u(x)=k^2u(x)
\label{math}
\eeq 
with
\beq
q(x)=\frac{m\ell^2}{\hbar^2}U(\ell x),\qquad\quad k=\ell\kappa
\eeq 
and has asymptotics
\beq
u_{\rm as}(x)={\rm e}^{-ikx}-S(k){\rm e}^{ikx},\qquad\quad 
{\cal T}(\kappa)=S(\kappa\ell).
\eeq 
The length scale is arbitrary but it is convenient to choose $\ell=2L$, where
$L$ is the Compton wavelength of the particle, $L=\hbar/mc$. With this choice
\beq
S_{\rm NR}(p_1,p_2)=-S\left(\frac{p_1-p_2}{mc}\right),\qquad\quad
U(x)=\frac{mc^2}{4}q\left(\frac{x}{2L}\right).
\eeq 

Our aim is to find a suitable effective potential $U(x)$ that, by solving the
corresponding nonrelativistic Schr\"odinger equation, leads to the physical, i.e.
relativistic, scattering S-matrix as function of the momentum of the particles.
Thus we require 
\beq
S\left(\frac{p_1-p_2}{mc}\right) \sim \Sigma\left(
{\rm arcsinh}\,\left(\frac{p_1}{mc}\right)
-{\rm arcsinh}\,\left(\frac{p_2}{mc}\right) \right).
\eeq 
Clearly, it is impossible to find such an effective potential in general, since the true
(relativistic) S-matrix is a function of the rapidity difference, whereas
the non-relativistic formula depends on the momentum difference. The identification 
is possible only approximately at low energies, where $p_j\approx mc\theta_j$.

There are, however, two important special cases, where exact identification is 
possible. In the laboratory (fixed target) frame of the scattering we can require
\beq
({\rm LAB})\ \ \qquad
S_I(k)=\Sigma\big({\rm arcsinh}\,(k)\big),\qquad\quad p_1=kmc,\quad p_2=0.
\label{LAB}
\eeq 
Similarly, in the centre of mass frame we require
\beq
({\rm COM})\qquad
S_{II}(k)=\Sigma\big(2\,{\rm arcsinh}\,(k/2)\big),\qquad p_1=-p_2=kmc/2.
\label{COM}
\eeq 
The resulting effective potentials $U_I(x)$ and $U_{II}(x)$ will be different.
The price we have to pay is frame dependence.

The problem we have to solve in both cases is to find the
potential $q(x)$ in (\ref{math}) if the corresponding S-matrix $S(k)$
is given. We are interested in potentials with a strong repulsive core,
which means that $q(x)$ has to be singular when the relative distance $x$
approaches zero. This leads us to the mathematical problem of quantum inverse
scattering with singular potentials, which is discussed in the next section.

%% file: singular
Quantum inverse scattering, the problem of finding the potential from 
scattering data, is a classical problem in quantum mechanics. It has been
completely solved in the one-dimensional case \cite{QIS1a,QIS1b,QIS1c} both
for the entire line and the half line cases. The latter case is more important
because the same mathematical problem emerges for three-dimensional
spherically symmetric potentials after partial wave expansion. Here we will 
also be interested in this case, because we consider strongly repulsive 
potentials. The details of the reconstruction procedure depend on the
class of the potentials and the simplest case is that of regular potentials
\cite{QIS2}. We will proceed along the lines presented in \cite{QIS2}, with 
some modifications necessary due to the singular core of our potentials.

We will consider the Schr\"odinger equation on the half line $x\geq0$
\beq
-u^{\prime\prime}(x)+q(x)u(x)=k^2u(x)
\label{diff}
\eeq
with boundary condition $u(0)=0$.
We will assume that the potential $q(x)$ is singular as $x\to0$, more precisely
we assume
\beq
q(x)\sim\frac{p(p-1)}{x^2},\qquad\quad x\to0,
\eeq
where $p>1$. (Later we will see that we recover the results for regular 
potentials in the limit $p\to1$.) We also assume that
\beq
q(x)\to0\qquad\quad{\rm as}\qquad\quad x\to\infty,
\eeq
and that it vanishes faster than $1/x^2$.

\subsection{Direct scattering}

For any given $k$, we will need three special solutions of the differential 
equation (\ref{diff}). The physical solution $\varphi(x,k)$ is defined by
its regular behaviour near the origin,
\beq
\varphi(x,k)=x^p[1+{\rm O}(x)],\qquad\quad x\to0.
\eeq
The singular solution $\tilde\varphi(x,k)$ is defined by the requirement
\beq
\tilde\varphi(x,k)=x^{1-p}[1+{\rm O}(x)],\qquad\quad x\to0.
\eeq
Finally the Jost solution is defined to have large $x$ asymptotics
\beq
f(x,k)={\rm e}^{ikx}[1+{\rm O}(1/x)],\qquad\quad x\to\infty.
\eeq
In addition to the scattering solutions with real momentum $k$, the
Schr\"odinger equation (\ref{diff}) may have normalizable bound state solutions
with imaginary $k$ (negative energy). Since in our main example in this paper
(soliton-soliton interaction in the Sine-Gordon model) there are no bound
states we will discuss here the case without bound states. It is easy to
work out the modifications necessary for potentials with bound states.

Since the second order differential equation (\ref{diff}) has only two 
linearly independent solutions, any of the above solutions can be expressed as
linear combinations of the other two. For example, the Jost solution
can be written as
\beq
f(x,k)=\tilde f(k)\varphi(x,k)+f(k)\tilde\varphi(x,k)
\label{lin1}
\eeq
with some coefficients $\tilde f(k)$, $f(k)$. $f(k)$ is called the Jost
function and plays an important role in scattering theory\footnote{For the
case of regular potentials $\varphi(0,k)=0$, $\tilde\varphi(0,k)=1$
and $f(k)$ is simply given by $f(0,k)$.}. It can be shown
that $f(k)$ can alternatively be defined by the linear combination
\beq
\varphi(x,k)=\frac{2p-1}{2ik}\left\{f(-k)f(x,k)-f(k)f(x,-k)\right\}.
\label{lin2}
\eeq
For real $k$
\beq
f^*(x,k)=f(x,-k)\qquad\qquad{\rm and}\qquad\qquad
f^*(k)=f(-k)
\eeq
and if we introduce the modulus and phase of $f(k)$ by writing
\beq
f(k)=\vert f(k)\vert{\rm e}^{-i\delta(k)}
\eeq
we see that
\beq
\vert f(k)\vert=\vert f(-k)\vert\qquad{\rm and}\qquad \delta(-k)=
-\delta(k)\quad{\rm mod\ }2\pi.
\eeq
From (\ref{lin2}) we see that  for large $x$ asymptotically
\beq
\varphi(x,k)\approx -\frac{2p-1}{2ik}f(k)\left\{{\rm e}^{-ikx}-S(k){\rm e}^{ikx}
\right\}=\frac{2p-1}{k}\vert f(k)\vert\sin[kx+\delta(k)].
\eeq
Here
\beq
S(k)=\frac{f(-k)}{f(k)}={\rm e}^{2i\delta(k)}
\eeq
and $\delta(k)$ is the phase shift.

It is possible to show that the large $k$ behaviour of the Jost function is
\beq
f(k)\approx\frac{\Gamma(2p-1)}{\Gamma(p)}(-2ik)^{1-p}[1+{\rm O}(1/k)].
\eeq
This gives
\beq
\delta(k)=\frac{\pi}{2}(1-p)+{\rm O}(1/k),\qquad\quad\delta(\infty)=
\frac{\pi}{2}(1-p).
\eeq
Since (mod $2\pi$) $\delta(-\infty)=-\delta(\infty)$, $S(\infty)$
and $S(-\infty)$ are not the same in general, except for integer $p$,
in which case
\beq
S(\infty)=S(-\infty)=(-1)^{p-1}.
\eeq

The physical solutions $\varphi(x,k)$ satisfy the completeness relation
\beq
\frac{2}{\pi}\int_0^\infty\frac{k^2{\rm d}k}{(2p-1)^2\vert f(k)\vert^2}
\varphi(x,k)\varphi(y,k)=\delta(x-y).
\label{complete}
\eeq

An other important object in inverse scattering theory is the transformation
kernel $A(x,y)$. It is defined as the unique solution of the Goursat problem
\beq
\frac{\partial^2}{\partial x^2}A(x,y)=
\frac{\partial^2}{\partial y^2}A(x,y)+q(x)A(x,y),
\eeq
\beq
-2\frac{{\rm d}}{{\rm d}x}A(x,x)=q(x),
\eeq
\beq
\lim_{(x+y)\to\infty}A(x,y)=
\lim_{(x+y)\to\infty}\frac{\partial}{\partial x}A(x,y)=
\lim_{(x+y)\to\infty}\frac{\partial}{\partial y}A(x,y)=0.
\eeq
This transformation kernel can be used to define the unitary operator
$\hat{\cal A}$ which maps the solutions of the free problem onto those
of the interacting problem with potential $q(x)$. The action of $\hat{\cal A}$
is defined by
\beq
\big(\hat{\cal A}{\cal F}\big)(x)={\cal F}(x)+\int_x^\infty{\rm d}y
A(x,y){\cal F}(y)
\eeq
and the mapping is
\beq
f_k=\hat{\cal A}E_k,\qquad\qquad f_k(x)=f(x,k),\qquad E_k(x)={\rm e}^{ikx}.
\eeq

\subsection{Inverse scattering}

Starting from the completeness relation (\ref{complete}), by acting on it
with the inverse of the unitary operator $\hat{\cal A}$, one can derive the
most important equation of inverse scattering, the Marchenko integral equation.
We have followed the steps presented in \cite{QIS2} for regular potentials.
In our case with singular potential one has to be careful because unlike
for regular potentials, $\delta(\infty)\not=0$ here. The result is that
$A(x,y)$ satisfies the Marchenko equation 
\beq
F(x+y)+A(x,y)+\int_x^\infty{\rm d}s\,A(x,s)\,F(s+y)=0,\qquad\quad y>x>0,
\label{Marchenko}
\eeq
where
\beq
F(x)=\frac{1}{2\pi ix}\int_{-\infty}^\infty{\rm d}k{\rm e}^{ikx}S^\prime(k).
\label{F1}
\eeq
The Marchenko equation (\ref{Marchenko}) is of the same form as for regular 
potentials, only the definition of $F(x)$ had to be modified. In the special 
case of integer $p$, an alternative form of (\ref{F1}) is obtained by 
partial integration
\beq
F(x)=\frac{1}{2\pi}\int_{-\infty}^\infty{\rm d}k{\rm e}^{ikx}
\left[(-1)^{p-1}-S(k)\right].
\label{F2}
\eeq
For $p=1$ the standard formula \cite{QIS2} is reproduced.

Quantum inverse scattering now proceeds in three steps. The first step is
to calculate $F(x)$ using the scattering data $S(k)$ in (\ref{F1}) or
(\ref{F2}). The second step is to solve (\ref{Marchenko}) for $A(x,y)$.
The third and final step is to use
\beq
-2\frac{{\rm d}}{{\rm d}x}A(x,x)=q(x)
\label{third}
\eeq
to determine $q(x)$.

%% file: LAB
In this section we carry out the three steps of quantum inverse scattering
to determine the effective SG potential that exactly reproduces
the SG soliton-soliton scattering in the laboratory frame (case I). The SG
S-matrix is given in Appendix B.

For simplicity, we deal with integer $p$ only.
Using the identification (\ref{LAB}) and the SG S-matrix (\ref{ss}) we have
\beq
S_I(k)=\prod_{m=1}^{p-1}\frac{s_m-ik}{s_m+ik},\qquad\quad s_m=\sin(\nu\pi m).
\label{SI}
\eeq

The first step is to calculate $F(x)$. For the above S-matrix (\ref{F2}) is 
easily evaluated with the help of the residue theorem and we obtain
\beq
F(x)=-\sum_{m=1}^{p-1}R_m\,{\rm e}^{-s_mx},\qquad\qquad
R_m=2s_m\prod_{n\not=m}\frac{s_n+s_m}{s_n-s_m}.
\label{FLAB}
\eeq

The next step is to solve the Marchenko equation for $A(x,y)$. For $F(x)$ 
given by (\ref{FLAB}) we have to solve
\beq
-\sum_m R_m{\rm e}^{-s_m(x+y)}+A(x,y)-\sum_m R_m{\rm e}^{-s_m y}
\int_x^\infty{\rm d}w\,A(x,w){\rm e}^{-s_m w}=0.
\label{mar1}
\eeq
We see that the $y$ dependence of $A(x,y)$ must be of the form
\beq
A(x,y)=\sum_m R_m a_m(x){\rm e}^{-s_my}.
\eeq
When this expression is substituted back to (\ref{mar1}) we find
\beq
a_m(x)={\rm e}^{-s_mx}+\sum_nR_n\int_x^\infty
{\rm d}w\, a_n(x){\rm e}^{-(s_m+s_n)w}.
\eeq
The $w$ integration can be performed and we get
\beq
a_m(x)={\rm e}^{-s_mx}+\sum_nR_n\,a_n(x)\frac{1}{s_m+s_n}{\rm e}^{-(s_m+s_n)x},
\eeq
which can be further simplified by introducing
\beq
a_m(x)={\rm e}^{-s_mx}\,b_m(x),\qquad\qquad z_m(x)=R_m\,{\rm e}^{-2s_mx}.
\eeq
We finally obtain the equations
\beq
b_m=1+\sum_n\frac{z_n\,b_n}{s_m+s_n}.
\label{algebra}
\eeq
This way the Marchenko integral equation is reduced to an algebraic problem.
We have to solve (\ref{algebra}) for the $b_m$ variables and using this 
solution we can write
\beq
A(x,y)=\sum_mR_m\,b_m(x)\,{\rm e}^{-s_m(x+y)}.
\eeq
Finally $A(x,x)$ is given by
\beq
A(x,x)=\sum_mb_m(x)\,z_m(x).
\label{bz}
\eeq

\input fig3

The solution of this algebraic problem turns out to be very simple. We can
rearrange (\ref{algebra}) to the matrix form
\beq
\sum_n{\cal M}_{mn}\,b_n=1
\label{mform1}
\eeq
where
\beq
{\cal M}_{mn}(x)=\delta_{mn}-\frac{z_n(x)}{s_m+s_n}.
\eeq
As shown in Appendix C, the solution is the logarithmic derivative of the
determinant of this matrix,
\beq
A(x,x)=\frac{{\rm d}}{{\rm d}x}\ln{\cal D}(x),\qquad\qquad {\cal D}(x)=
{\rm Det}\,\left({\cal M}(x)\right).
\label{Axx}
\eeq
The final results can be further simplified if we introduce the \lq\lq reduced"
determinant $\hat{\cal D}$ by writing
\beq
{\cal D}=2^{p-1}\,\left(\prod_{k<l}\frac{1}{s_k-s_l}\right)
\left(\prod_m{\rm e}^{-s_mx}\right)\hat{\cal D}.
\eeq
Since the determinant is a totally symmetric expression of the variables $s_j$
and the prefactor is totally antisymmetric, the reduced determinant must 
also be totally antisymmetric. Moreover, it turns out to be a polynomial
in the variables $s_j$, $H_j$, $C_j$, where
\beq
H_j=\sinh(s_jx),\qquad\qquad C_j=s_j \cosh(s_jx).
\eeq

It is easy to see that for $p=2$ we have $\hat{\cal D}=H_1$. We have calculated
the reduced determinant for $p=3,4,5$ using Mathematica. For $p=3$
\beq
\hat{\cal D}=C_1H_2-C_2H_1,
\eeq
for $p=4$
\beq
\hat{\cal D}=-(s_1^2-s_2^2)C_3H_1H_2
+(s_1^2-s_3^2)C_2H_1H_3-(s_2^2-s_3^2)C_1H_2H_3,
\eeq
finally for $p=5$ Mathematica found
\beq
\hat{\cal D}=-(s_1^2-s_2^2)(s_3^2-s_4^2)C_1C_2H_3H_4\quad+\quad
{\rm 5\ anti-permutations}.
\label{p5}
\eeq
The 5 additional terms on the right hand side of (\ref{p5}) 
make it totally antisymmetric.

From the above formulas it is clear how $\hat{\cal D}$ can be constructed
from the variables $s_j$, $H_j$, $C_j$ in general. Since our calculation is
algebraic, it must be valid also for the case discussed in Appendix A, since
the corresponding S-matrix is also of the form (\ref{SI}), with $s_m=m$.
It is a very non-trivial check on our result that in this case (\ref{Axx})
is reduced to 
\beq
\frac{p(p-1)}{2}[\coth(x)-1],
\eeq
which is by far not obvious, but turns out to be true.

The small $x$ expansion of (\ref{Axx}) takes the form
\beq
\begin{split}
A(x,x)&=\frac{p(p-1)}{2x}-\sum_js_j+\frac{x}{2p-1}\left(\sum_j s_j^2\right)
+{\rm O}(x^3)\\
&=\frac{p(p-1)}{2x}-\frac{1}{2}\cot\frac{\pi\nu}{2}+\frac{x}{4}
+{\rm O}(x^3).
\end{split}
\eeq
The strength of the $x\to0$ singularity is exactly the same as we assumed
at the beginning of our considerations. 

We have compared the (integrated) laboratory frame effective potential 
and the (integrated) zero-momentum potential in Figs. 1,2
for $p=3,4$.

\input fig4

\clearpage

%% file: fig3


\begin{figure}[htb]
\begin{flushleft}
\hskip 15mm
\leavevmode
\centerline{\includegraphics[width=12cm]{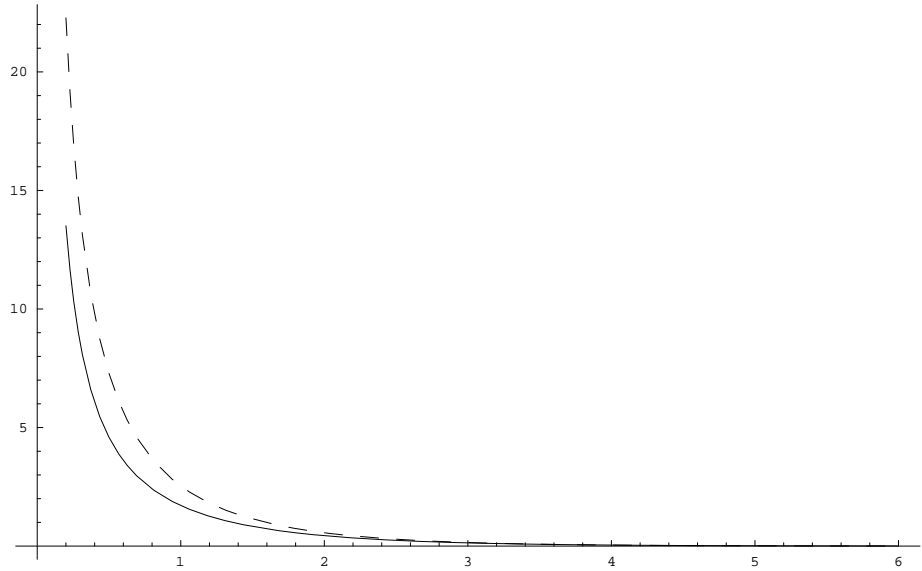} }
\end{flushleft}
\vspace{-3.5cm}
\caption{{\footnotesize
Comparison of the integrated effective potential $A(x,x)$ (solid) and the
corresponding  zero-momentum $A_o(x,x)$ (dashed) for $p=3$.
}}
\label{LAB3}
\end{figure}


%% file: fig4


\begin{figure}[htb]
\begin{flushleft}
\hskip 15mm
\leavevmode
\centerline{\includegraphics[width=12cm]{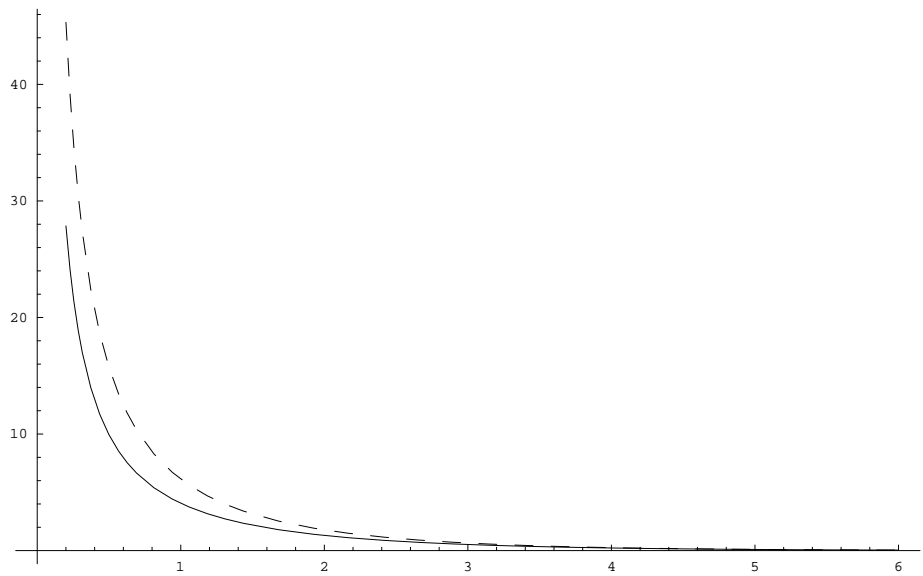} }
\end{flushleft}
\vspace{-3.5cm}
\caption{{\footnotesize
Comparison of the integrated effective potential $A(x,x)$ (solid) and the
corresponding  zero-momentum $A_o(x,x)$ (dashed) for $p=4$.
}}
\label{LAB4}
\end{figure}


%% file: COM
In this section we calculate the SG effective potential in the centre of mass
frame. Again, we restrict our attention to integer $p$.
Using (\ref{COM}) and (\ref{ss}) we have
\beq
S_{II}(k)=\prod_{m=1}^{p-1}\frac{s_m-ik\rho(k)}{s_m+ik\rho(k)},
\qquad\quad \rho(k)=\sqrt{1+\frac{k^2}{4}}.
\eeq
This can be equivalently written
\beq
S_{II}(k)=(-1)^{p-1}+\sum_m\frac{R_m}{s_m+ik\rho(k)}
\eeq
and correspondingly, using (\ref{F2}),
\beq
F_{II}(x)=-\sum_mR_m{\cal F}(x;s_m),
\eeq
where
\beq
{\cal F}(x;\sigma)=\frac{1}{2\pi}\int_{-\infty}^\infty{\rm d}k\,
\frac{{\rm e}^{ikx}}{\sigma+ik\rho(k)}.
\label{calF}
\eeq
Let us introduce the notations
\beq
\sigma=\sin\varphi,\qquad \tilde\sigma=\cos\varphi,\qquad\qquad
\alpha=\sin\frac{\varphi}{2},\qquad \beta=\cos\frac{\varphi}{2}.
\eeq
The integrand of (\ref{calF}) in the upper half plane has poles at
$k=2i\alpha$, $2i\beta$ with residues $-i\beta/\tilde\sigma$,
$i\alpha/\tilde\sigma$ respectively and a cut starting at $k=2i$ and
going up along the imaginary axis. We can evaluate the Fourier integral
by closing the contour with a half-circle at infinity and using the 
residue theorem, but we have to add the contribution of the cut as well.
The contribution of the poles is
\beq
{\cal F}^{{\rm pole}}(x;\sigma)=\frac{1}{\tilde\sigma}\left(
\beta{\rm e}^{-2\alpha x}-\alpha{\rm e}^{-2\beta x}\right)
\eeq
and we can write
\beq
{\cal F}(x;\sigma)={\cal F}^{{\rm pole}}(x;\sigma)+
{\cal F}^{{\rm cut}}(x;\sigma),
\eeq
where
\beq
{\cal F}^{{\rm cut}}(x;\sigma)=-\frac{1}{\pi}\int_2^\infty
{\rm d}\kappa\,\frac{\kappa R{\rm e}^{-\kappa x}}{\sigma^2+\kappa^2 R^2},
\qquad\quad
R=\sqrt{\frac{\kappa^2}{4}-1}.
\eeq
This form is more suitable for numerical evaluation because instead of
an oscillating integrand it contains a decaying exponential.

\input COM2

We calculated $F_{II}(x)$ numerically for $p=2$, $3$ and by discretizing the
integrals solved the corresponding Marchenko equations numerically.
The results are shown in Figs. 3,4. For comparison we also show in these
plots the corresponding LAB frame (integrated) effective potentials. 
It can be seen that the frame dependence is weak: both effective potentials 
have the same qualitative features and are close to each other. 
The expected $1/x$ short distance behaviour is also reproduced.
We can conclude that the notion of effective potential makes sense in this
model.

\input COM3


\clearpage

%% file: COM2


\begin{figure}[htb]
\begin{flushleft}
\hskip 15mm
\leavevmode
\centerline{\includegraphics[width=12cm]{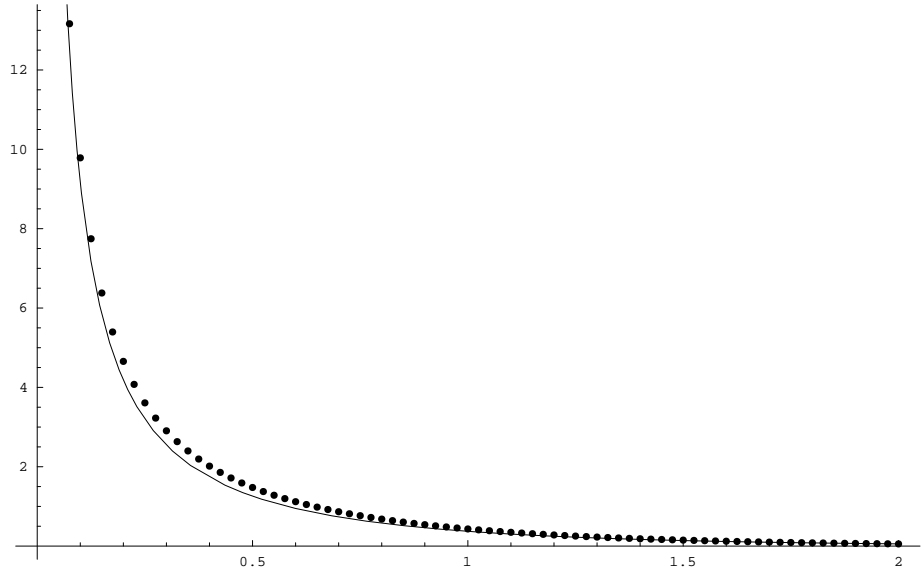} }
\end{flushleft}
\vspace{-3.5cm}
\caption{{\footnotesize
The integrated effective potential in the COM frame for $p=2$ (dots).
For comparison the analytically obtained LAB frame integrated effective 
potential $A(x,x)$ (solid) is also shown.
}}
\label{COM2}
\end{figure}


%% file: COM3


\begin{figure}[htb]
\begin{flushleft}
\hskip 15mm
\leavevmode
\centerline{\includegraphics[width=12cm]{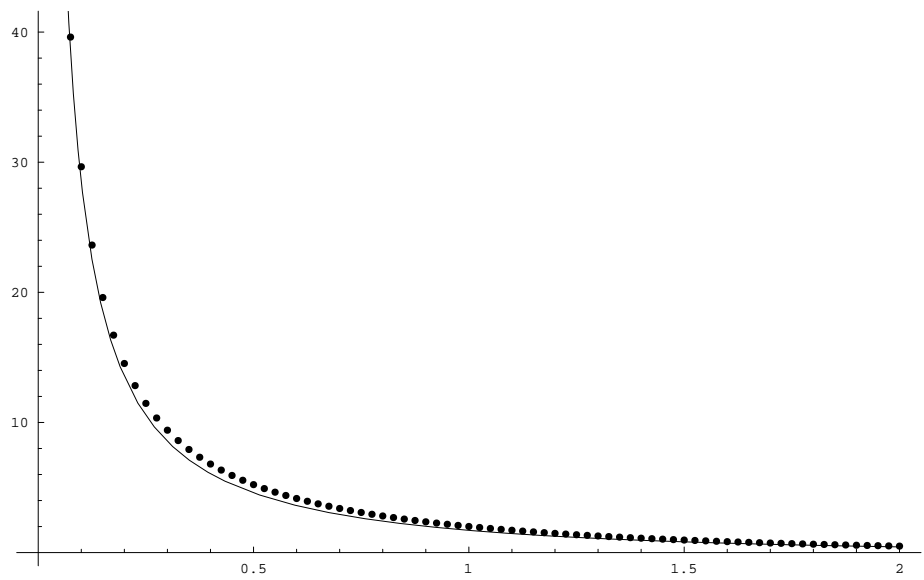} }
\end{flushleft}
\vspace{-3.5cm}
\caption{{\footnotesize
The integrated effective potential in the COM frame for $p=3$ (dots).
For comparison the analytically obtained LAB frame integrated effective 
potential $A(x,x)$ (solid) is also shown.
}}
\label{COM3}
\end{figure}


%% file: QISconcl
The phenomenological potential in nuclear phsysics has a limited range
of applicability because the very notion of a potential used in the
Schr\"odinger equation is a nonrelativistic concept which is meaningful
and valid (approximately) only below the $\pi$-production threshold.
The NBS potential as measured by the original HAL QCD method
\cite{Ishii:2006ec} is energy dependent (although this energy dependence is
moderate at low energies). An alternative possibility is to define
\cite{Aoki:2008hh,Aoki:2009ji} an energy-independent, 
but nonlocal \lq\lq potential".

$1+1$ dimensional integrable models are useful because the analogous
problems can be studied more explicitly. Moreover, since there is no
particle production in integrable models, the two-particle description
remains valid at all energies. It is possible to define an effective 
potential, which is energy independent and reproduces the scattering data
exactly. The price one has to pay for energy independence is that due to
the relativistic nature of the problem this effective potential becomes
frame dependent.

\input SUM

In this paper we studied the effective potential in the SG model. We 
calculated the effective potential algebraically in the laboratory frame
and numerically in the centre of mass frame using inverse scattering
techniques. Our results are summarized in Fig. \ref{SUM}, where the LAB and COM
frame effective potentials are compared and the zero-momentum potential
(obtained from the equivalent Ruijsenaars-Schneider formulation of the
model) is also shown. The three potentials are qualitatively very
similar and also close numerically. Our conclusion is that (at least in
this $1+1$ dimensional toy model) in spite of the problems discussed above
the effective potential remains a useful concept.


%% file: SUM


\begin{figure}[htb]
\begin{flushleft}
\hskip 15mm
\leavevmode
\centerline{\includegraphics[width=13cm]{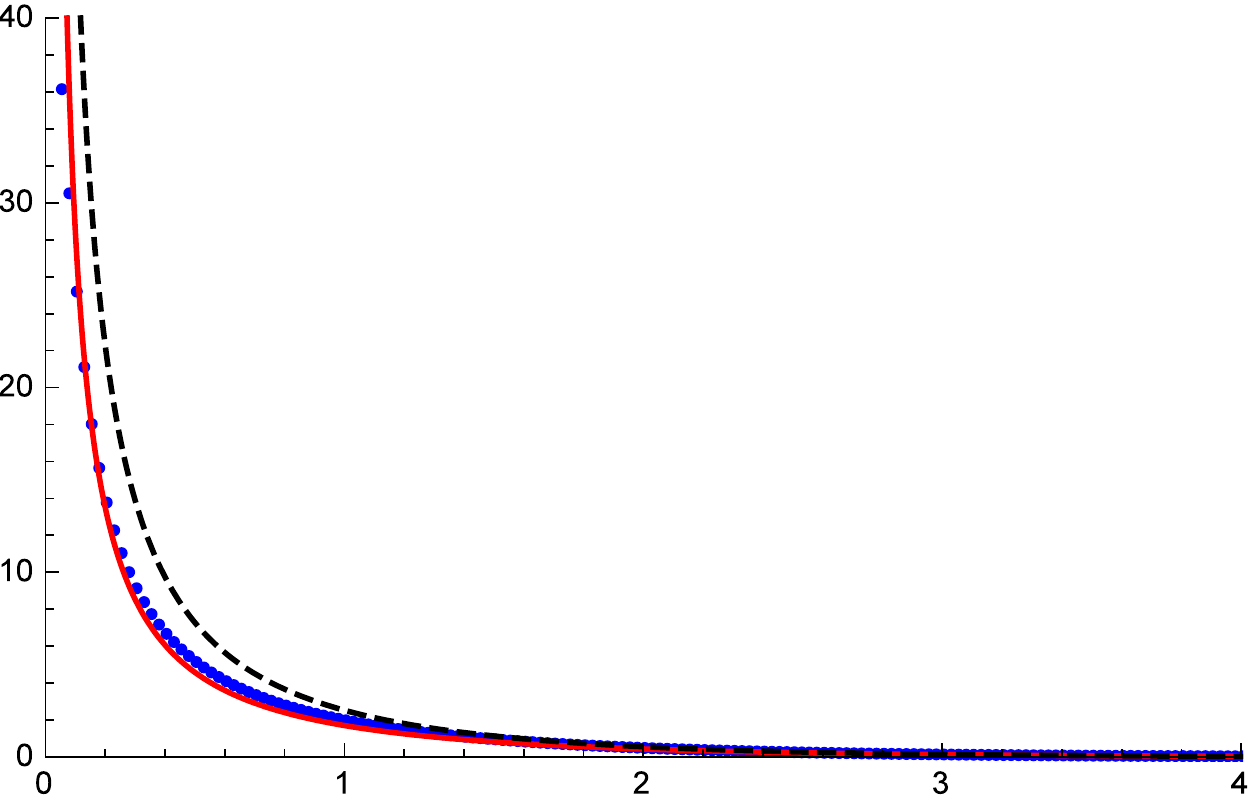}}
\end{flushleft}
\caption{{\footnotesize
Comparison of integrated SG effective potentials for $p=3$. The
solid (red) line, the (blue) dots and the dashed (black) line are
the LAB frame, the COM frame and the zero-momentum potential,
respectively.
}}
\label{SUM}
\end{figure}


%% file: sinh
To illustrate the steps of direct and inverse scattering, we take the
solvable potential
\beq
q(x)=\frac{p(p-1)}{\sinh^2x}.
\label{sinh}
\eeq
The solution of the Schr\"odinger equation (\ref{diff}) with this potential
is well known and proceeds by introducing the new variables
\beq
u(x)={\rm e}^{ikx}F(z), \qquad\qquad z=\frac{1}{2}(1+\coth x).
\eeq
The Schr\"odinger equation becomes
\beq
z(1-z)F^{\prime\prime}(z)+(1+ik-2z)F^\prime(z)+p(p-1)F(z)=0,
\eeq
which is the hypergeometric differential equation with parameters
\beq
a=p,\qquad\quad b=1-p,\qquad\quad c=1+ik.
\eeq
The hypergeometric differential equation has many solutions expressible
by Gauss' hypergeometric function ${}_2F_1$. The solutions we need are
\beq
\varphi(x,k)=\frac{1}{2^p}\left(1-{\rm e}^{-2x}\right)^p{\rm e}^{ikx}
{}_2F_1\left(p,p-ik,2p;1-{\rm e}^{-2x}\right),
\eeq
\beq
\tilde\varphi(x,k)=2^{p-1}\left(1-{\rm e}^{-2x}\right)^{1-p}{\rm e}^{ikx}
{}_2F_1\left(1-p-ik,1-p,2-2p;1-{\rm e}^{-2x}\right),
\eeq
\beq
f(x,k)=\left(1-{\rm e}^{-2x}\right)^p{\rm e}^{ikx}
{}_2F_1\left(p,p-ik,1-ik;{\rm e}^{-2x}\right).
\eeq
$\varphi(x,k)$ and $f(x,k)$ are always well defined by the above formula,
but the above expression for $\tilde\varphi(x,k)$ is valid only if $p$ is
not an integer or half-integer. This is a technical difficulty only and does 
not imply that $\tilde\varphi(x,k)$ does not exist in these cases. It only 
means that it cannot be simply expressed in terms of ${}_2F_1$. Moreover,
our formulas for $f(k)$ and the S-matrix are continuous and turn out to be
valid for integer/half-integer $p$ as well.

Using the well-known linear relations between the hypergeometric functions
of argument $z$ and argument $1-z$ we can read off the coefficients defined
by (\ref{lin1}). In this example they turn out to be
\beq
f(k)=\frac{1}{2^{p-1}}\,\frac{\Gamma(1-ik)\Gamma(2p-1)}
{\Gamma(p)\Gamma(p-ik)},
\eeq
\beq
\tilde f(k)=2^p\,\frac{\Gamma(1-ik)\Gamma(1-2p)}
{\Gamma(1-p-ik)\Gamma(1-p)}.
\eeq
It can be checked that using (\ref{lin2}) leads to the same expression
for $f(k)$.

The S-matrix is
\beq
S(k)=\frac{\Gamma(1+ik)\Gamma(p-ik)}
{\Gamma(1-ik)\Gamma(p+ik)}.
\eeq
As mentioned before, this derivation is not valid for integer $p$. 
Nevertheless, the formula  for the S-matrix remains valid for integer $p$ too.
Moreover, for integer $p$ it simplifies to
\beq
S(k)=\prod_{j=1}^{p-1}\frac{j-ik}{j+ik}.
\eeq

The simplest nontrivial case is $p=2$. The corresponding S-matrix is
\beq
S(k)=\frac{1-ik}{1+ik}
\eeq
and (\ref{F2}) gives
\beq
F(x)=-2{\rm e}^{-x}.
\eeq
For this $F(x)$ the Marchenko equation is easily solved and one finds
\beq
A(x,y)=\frac{{\rm e}^{-y}}{\sinh x}.
\eeq
Thus
\beq
A(x,x)=\coth x-1
\eeq
and using (\ref{third}) the potential (\ref{sinh}) is reproduced, as it should.

%% file: SGSmatrix
The Sine-Gordon (SG) model is perhaps the most studied two-dimensional 
integrable field theory. Its spectrum and S-matrix is exactly known from
its bootstrap solution \cite{ZZ}. Moreover, an equivalent relativistic
quantum mechanical description exists, the Ruijsenaars-Schneider
model \cite{RS1,RS2}.

The SG field theory Lagrangian is\footnote{Here we use the $\hbar=c=1$ 
system of units as usual in relativistic quantum field theory.} 
\beq
{\cal L}=\frac{1}{2}\left(\dot\phi^2-\phi^{\prime2}\right)+
\frac{\mu^2}{\beta^2}\cos(\beta\phi),
\eeq
where $\mu$ is a mass parameter and $\beta$ is the SG coupling. The model 
is well-defined only if $0<\beta^2<8\pi$. $\beta^2=4\pi$ is the free 
fermion point. We will use the parameters 
\beq
p=\frac{4\pi}{\beta^2}>\frac{1}{2}\qquad\qquad{\rm and}\qquad\qquad
\nu=\frac{1}{2p-1}.
\eeq
The spectrum of the model includes a U$(1)$ doublet of particles (soliton
and antisoliton of mass $m$). There are also soliton-antisoliton bound states
(breathers), whose mass spectrum is given by
\beq
m_k=2m\sin\left(\frac{\pi\nu k}{2}\right),\qquad\quad k=1,2,\dots <2p-1.
\eeq
The soliton mass is related to the Lagrangian mass parameter by
\beq
m=\frac{2p-1}{\pi}\mu.
\eeq

The full S-matrix of the model (scattering among solitons, antisolitons,
breathers) is completely known \cite{ZZ}, but in this paper we only need
the soliton-soliton scattering S-matrix. Here there are no bound states
and it is given by the formula
\beq
\Sigma(\theta)=\exp\left\{i\int_0^\infty\frac{{\rm d}\omega}{\omega}
\sin\left(\frac{2\theta\omega}{\pi}\right)\frac{\sinh\big((\nu-1)\omega\big)}
{\cosh(\omega)\sinh(\nu\omega)}\right\}.
\eeq
Analitically continuing $\Sigma(\theta)$ to the complex rapidity strip
$0<{\rm Im}\,\theta<\pi$ we find that it has poles at
\beq
\theta_k=i\pi k\nu,\qquad\qquad k=1,2,\dots<2p-1.
\eeq
In the large rapidity limit
\beq
\Sigma(\pm\infty)={\rm e}^{\pm i\pi(1-p)}.
\eeq
$p$ is a continuous parameter, but the S-matrix simplifies for integer $p$.
In this case $\Sigma(\theta)$ is a function of $\sinh(\theta)$ and is given by
\beq
\Sigma(\theta)=\prod_{j=1}^{p-1}\frac{s_j-i\sinh(\theta)}{s_j+i\sinh(\theta)},
\label{ss}
\eeq
where
\beq
s_j=\sin(\nu\pi j),\qquad\qquad j=1,2,\dots,p-1.
\eeq

The Ruijsenaars-Schneider (RS) model \cite{RS1,RS2} is an integrable 
relativistic quantum mechanical model whose dynamics and S-matrix is completely
equivalent to that of the SG field theory. From the RS description it is 
possible to read off the corresponding zero-momentum potential \cite{RS2,BZ}.
In our conventions it reads (after restoring the constants $\hbar$, $c$)
\beq
U_o(x)=\frac{mc^2}{\sinh^2\left(\frac{\pi\nu x}{2L}\right)}.
\eeq
After rescaling by $\ell=2L$ we get
\beq
q_o(x)=\frac{4}{\sinh^2(\pi\nu x)}.
\eeq
Although it has no special meaning in the SG context, for later convenience
we introduce
\beq
A_o(x,x)=\frac{2}{\pi\nu}\left[\coth(\pi\nu x)-1\right].
\eeq
Its relation to $q_o(x)$ is analogous to (\ref{third}).

%% file: detsol
\newcommand{\half}{{\frac{1}{2}}}
\def\2{{\half}}
\newcommand{\const}{\mathop{\rm const}\nolimits}
\def\parag{\hfil\break} 
\def\kikezd{\parag\underbar}
\def\bA{{\bm{A}}}
\def\su{{\rm su}}
\def\p{{\partial}}
\def\bk{{\bf k}}
\def\bR{{\mathds{R}}}
\def\bbeta{{\bm{\beta}}}
\def\tbbeta{{\widetilde{\bm{\beta}}}}
\def\bgamma{{\bm{\gamma}}}
\def\bGamma{{\bm{\Gamma}}}
\def\bomega{{\bm{\omega}}}
\def\bSigma{{\bm{\Sigma}}}
\def\bp{{\bm{p}}}
\def\tbp{\tilde{\bm{p}}}
\def\tbx{\tilde{\bm{x}}}
\def\tV{\widetilde{V}}
\def\tf{\tilde{f}}
\def\tg{\tilde{g}}
\def\tY{\tilde{Y}}
\def\ba{{\bm{a}}}
\def\bc{{\bf c}}
\def\bu{{\bm{u}}}
\def\bOmega{\mbox{\boldmath$\omega$}}
\def\bnabla{\mbox{\boldmath$\nabla$}}
\def\br{{\bm{r}}}
\def\bL{{\bm{L}}}
\def\bg{{\bm{g}}}
\def\bQ{{\bm{Q}}}
\def\bP{{\bm{P}}}
\def\bE{{\bm{E}}}
\def\bB{{\bm{B}}}
\def\bb{{\bm{b}}}
\def\bD{{\bm{D}}}
\def\bnabla{{\bm{\nabla}}}
\def\bq{{\bm{q}}}
\def\bp{{\bm{p}}}
\def\bS{{\bm{S}}}
\def\bs{{\bm{s}}}
\def\bv{{\bm{v}}}
\def\hbp{{\widehat{\bm{p}}}}
\def\hbb{{\widehat{\bm{b}}}}
\def\bx{{\bm{x}}}
\def\bz{{\bm{z}}}
\def\by{{\bm{y}}}
\def\bW{{\bm{W}}}
\def\wbx{{\widetilde{\bm{x}}}}
\def\wbp{{\widetilde{\bm{p}}}}
\def\bX{{\bm{X}}}
\def\bY{{\bm{Y}}}
\newcommand{\np}{\vert\bp\vert}
\newcommand{\nq}{\vert\bq\vert}
\def\beqa{\begin{eqnarray}}
\def\eeqa{\end{eqnarray}}
\def\nn{\nonumber}
\def\barray{\left(\begin{array}}
\def\earray{\end{array}\right)}
\def\barraynb{\begin{array}}
\def\earraynb{\end{array}}
\def\IR{{\mathds{R}}} 
\def\IZ{{\mathds{Z}}}
\def\IC{{\mathds{C}}} 
\def\so{{\rm so}}
\def\SO{{\rm SO}}
\def\ort{{\mathfrak{o}}}
\def\Ort{{\rm O}}
\def\smallover#1/#2{\hbox{$\textstyle\frac{#1}{#2}$}} %
\def\vx{{\vec{x}}}
\def\vp{{\vec{p}}}
\def\vy{{\vec{y}}}
\def\vA{{\vec{A}}}
\def\vB{{\vec{B}}}
\def\vE{{\vec{E}}}
\def\va{{\vec{a}}}
\def\vnabla{{\overrightarrow{\nabla}}}
\def\vbeta{{\vec{\beta}}}
\newcommand{\Sp}{\mathrm{Sp}}
\newcommand{\cJ}{\mathcal{J}}
\newcommand{\sfP}{\mathsf{P}}
\newcommand{\sfp}{\mathsf{p}}
\newcommand{\cP}{\mathcal{P}}
\newcommand{\hp}{\widehat{p}}
\newcommand{\sfs}{\mathsf{s}}
\newcommand{\homega}{{\widehat{\omega}}}
\newcommand{\bbR}{\mathbb{R}}
\newcommand{\la}{{\langle}}
\newcommand{\ra}{{\rangle}}
\newcommand{\fg}{{\mathfrak{{g}}}}
\newcommand{\cO}{{\mathcal{O}}}
\newcommand{\cE}{{\mathcal{E}}}
\newcommand{\Tr}{\mathrm{Tr}}
\newcommand{\rg}{\mathrm{g}}
\newcommand{\belle}{\boldsymbol{\ell}}
\newcommand{\se}{\mathrm{e}}
\newcommand{\rE}{\mathrm{E}}
\newcommand{\SE}{\mathrm{SE}}
\newcommand{\sfA}{\mathcal{A}}

\newcommand{\Ad}{{\mathrm{Ad}}}
\newcommand{\Coad}{{\mathrm{Coad}}}

Let us recall (\ref{mform1}), the set of equations we have to solve for $b_m$
written in matrix form.
\beq
\label{mform}
\sum_n{\cal M}_{mn}b_n=e_m,\qquad\quad e_m=1,\quad m=1,2,\dots p-1,
\eeq
where
\beq
{\cal M}_{mn}=\delta_{mn}-\frac{z_n}{s_n+s_m}.
\label{calM}
\eeq
The solution can be written in matrix language as
\beq
b_m=\sum_n\left({\cal M}^{-1}\right)_{mn}e_n=
\sum_n\left({\cal M}^{-1}\right)_{mn}
\eeq
and the integrated potential, which is given by (\ref{bz}), as
\beq
A(x,x)=\sum_{m,n}z_m\left({\cal M}^{-1}\right)_{mn}.
\eeq
Let us denote the determinant of (\ref{calM}) by ${\cal D}$,
\beq
{\cal D}={\rm Det}({\cal M})
\eeq
and its logarithmic derivative by
\beq
{\cal A}(x)=\frac{{\rm d}}{{\rm d} x}\ln{\cal D}.
\label{calA}
\eeq
We conjecture that
\beq
\label{conjecture}
A(x,x)={\cal A}(x).
\eeq
For (\ref{calA}) an alternative expression is
\beq
{\cal A}(x)=\frac{{\rm d}}{{\rm d} x}\ln{\rm Det}\,({\cal M})=\Tr\left\{{\cal M}^{-1}
\frac{{\rm d}{\cal M}}{{\rm d} x}\right\}=
\sum_{m,n}\left({\cal M}^{-1}\right)_{mn}\frac{{\rm d}}{{\rm d} x}{\cal M}_{nm}.
\eeq
Since
\beq
\frac{{\rm d}}{{\rm d} x}{\cal M}_{mn}=\frac{2s_nz_n}{s_m+s_n},
\eeq
we can write
\beq
{\cal A}(x)=\sum_{m,n}z_m\left({\cal M}^{-1}\right)_{mn}\frac{2s_m}{s_m+s_n}=
\sum_{m,n}z_m\left({\cal M}^{-1}\right)_{mn}\frac{(s_m+s_n)+(s_m-s_n)}{s_m+s_n}.
\eeq
and further
\beq
{\cal A}(x)=A(x,x)+{\cal B}(x),
\eeq
where
\beq
{\cal B}(x)=
\sum_{m,n}z_m\left({\cal M}^{-1}\right)_{mn}\frac{s_m-s_n}{s_m+s_n}.
\eeq
Next we write the matrix ${\cal M}$ as a matrix product of a symmetric and a diagonal matrix:
\beq
{\cal M}={\cal K}\Delta,
\eeq
where
\beq
\Delta_{mn}=z_m\delta_{mn},\qquad\quad
{\cal K}_{mn}=\frac{1}{z_m}\delta_{mn}-\frac{1}{s_m+s_n},\qquad {\cal K}_{mn}={\cal K}_{nm}.
\eeq
The inverse in matrix form is
\beq
{\cal M}^{-1}=\Delta^{-1}{\cal K}^{-1}
\eeq
and in components
\beq
\left({\cal M}^{-1}\right)_{mn}=\frac{1}{z_m}\left({\cal K}^{-1}\right)_{mn}.
\eeq
So finally we have
\beq
{\cal B}(x)=
\sum_{m,n}\left({\cal K}^{-1}\right)_{mn}\frac{s_m-s_n}{s_m+s_n}=0,
\eeq
due to the symmetry of the inverse matrix ${\cal K}^{-1}$. This proves the conjecture.

%% file: bibliography.tex

  


\vfill\eject